\documentclass[aps]{revtex4}
\usepackage{amsmath,amssymb,amsfonts}
\usepackage{graphics,graphicx}

\begin{document}

\begin{abstract}
Two dimensional suspensions of spherical colloids subject to periodic
external fields exhibit a rich variety of molecular crystalline
phases. We study in simulations the ground state configurations of
dimeric and trimeric systems, that are realized on square and
triangular lattices, when either two or three macroions are trapped in
each external potential minimum. Bipartite orders of the checkerboard
or stripe types are reported together with more complex quadripartite
orderings, and the shortcomings of envisioning the colloids gathered
in a single potential minimum as a composite rigid object are
discussed. This work also sheds light on simplifying assumptions
underlying previous theoretical treatments and that made possible the
mapping onto spin models.
\end{abstract}

\title{Ground states of colloidal molecular crystals on periodic substrates}
\author{Samir El Shawish}
 \email{samir.elshawish@ijs.si}
\affiliation{Department of Theoretical Physics, Jo\v zef Stefan Institute, Jamova 39, 1000 Ljubljana, Slovenia}
\author{Jure Dobnikar}
 \affiliation{Department of Theoretical Physics, Jo\v zef Stefan Institute, Jamova 39, 1000 Ljubljana, Slovenia}
\author{Emmanuel Trizac}
 \affiliation{Universit\'e Paris-Sud, Laboratoire de Physique Th\'eorique et Mod\`eles Statistiques (CNRS UMR 8627), 91405 Orsay Cedex (France)}
 
\date{\today}
\maketitle

\section{Introduction}

Colloidal suspensions provide valuable systems for the study of
collective effects and phase transitions (see e.g \cite{Hunter} or
more specifically \cite{JPCM2} and references therein). Recent
advances in optical trapping techniques have enlarged the whole field
and opened the possibility to obtain quasi two-dimensional systems,
which can furthermore be subject to an external potential.  A
possibility to realize such external perturbations is through the
interference of laser beams, the typically micron-sized colloids being
attracted to the regions of highest intensity: for instance, unusual
phases have been reported in the case of 1D troughs \cite{Bech2001},
such as floating solids or locked smectic phases
\cite{Radzi,BechFrey}.

\begin{figure}[!h]
\includegraphics[width=0.6\textwidth]{fig1.eps}
\includegraphics[width=0.3\textwidth]{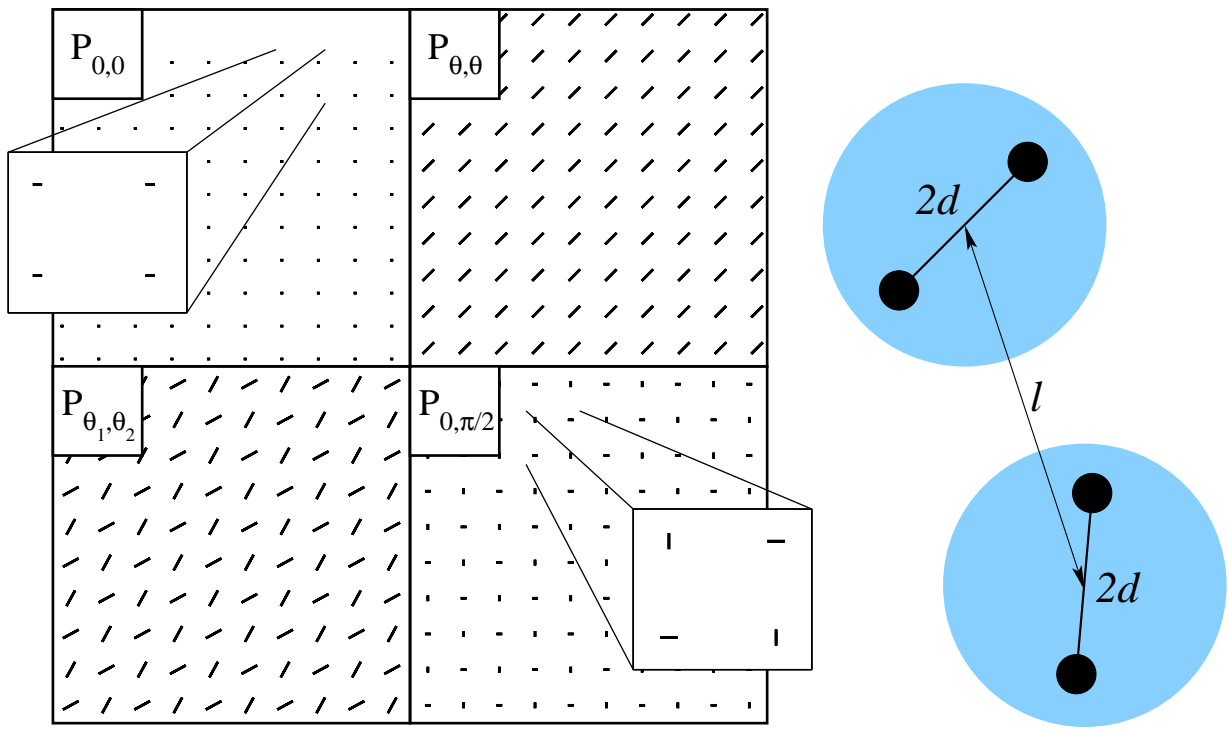}
\caption{ 
a) Several representative orientational structures observed on a 2D
rectangular lattice, for a stoichiometry $n=2$ (dimers). For
visualisation purposes, the two colloids forming a dimer are hereafter
linked by a thick line. The parameters associated with the different
phases are the following: for the structure denoted $P_{0,0}$
(ferromagnetic): $\kappa d=0.25$, $\kappa l=6$, $\alpha=0.9$;
$P_{\theta,\theta}$ (tilted ferromagnetic): $\kappa d=1$, $\kappa
l=4.5$, $\alpha=1$; $P_{\theta_1,\theta_2}$: $\kappa d=1.5$, $\kappa
l=6$, $\alpha=1$; and for the $P_{0,\pi/2}$ (antiferromagnetic) phase:
$\kappa d=0.5$, $\kappa l=6$, $\alpha=1$.  $1/\kappa$ is the inverse
Debye length and represents the range of the electrostatic screened
interactions between colloids. b) Schematic drawing showing dimers in
the traps (shaded circles). $l$ is the distance between the
neighbouring traps and $2d$ is the size of the "colloidal molecule". }
\label{fig:example}
\end{figure}

The situation of a two-dimensional periodic substrate, some instances
of which can be realized by 3 laser beams, is remarkably rich and has
been addressed experimentally \cite{Bech2002}, numerically
\cite{RO,half-integer,RO2005} and analytically \cite{Agra,Sarlah}. The
control parameters governing the static behaviour are numerous :
filling fraction (mean number of colloids per substrate minimum),
pinning amplitude (trap strength, increasing with laser intensity),
temperature, and concerning the substrate geometry, lattice spacing
and aspect ratio (i.e rectangular unit cell instead of a square
one). In addition, the colloids considered are highly charged objects,
so that the Debye length of the suspension, modified by changing the
salt content, is a crucial parameter. In the present paper, we
concentrate on the experimentally relevant ground state of those 2D
systems, where the long range orientational order observed has been
coined ``colloidal molecular crystal''. If the external laser
potential is strong enough, the colloids are irreversibly bound to the
potential minima. In cases when more than one colloid is trapped in a
single minimum we speak of a ''colloidal molecule''. Its size is
determined by the interplay between light forces and interparticle
repulsion. Figure \ref{fig:example}-a) provides an illustration of
typical phases observed when the ratio of the total number of colloids
in the system to the total number of external potential minima is
exactly two. Whereas defects are present at finite temperature, we
observe that the ground state is free of such objects, so that each
trap captures exactly two colloids.

The purpose of the present paper is twofold. The first goal is to
provide a thorough numerical investigation of orientational ordering
in colloidal molecular crystals on two-dimensional periodic
substrates. We will focus on stoichiometries 2 and 3, where dimers and
trimers are formed, respectively, with the underlying square or
triangular symmetry for the potential.  We will also address the
situations where the corresponding lattice unit cell is distorted by
changing the aspect ratio $\alpha$. The second objective is to
critically test several assumptions that led to the theoretical
frameworks used in Refs. \cite{Agra,Sarlah} to study such problems. In
section \ref{sec:rigid}, we will start with an approach, common to
Refs \cite{Agra,Sarlah}, where the composite objects ($n$-mers) formed
in each trap are considered as rigid entities with an orientational
degree of freedom only, while the confining potential is taken into
account implicitly. This allows for a significant reduction of the
complexity of the problem. Reference phase behaviour is thus obtained,
which will be tested against more realistic simulations in Section
\ref{sec:flexible}, where each colloid is resolved and the confining
potential explicitly accounted for.  Finally, we will also test the
relevance of the various effective potentials used in \cite{Agra} to
construct a tractable Hamiltonian allowing for analytical
progress. Our main finding and the ensuing consequences will be
summarised in section
\ref{sec:concl}.

%%%%%%%%%%%%%%%%%%%%%%%%%%%%%%%%%%%%%%%%%%%%%%%%%%%%%%%%%%%%%%%%%%%%%%%%%%%%%%%%%%%%%%%%%%%
\section{The rigid $n$-mer approach}
\label{sec:rigid}

The integer number $n$ of colloids that gather in a light potential
minimum \cite{rque} are subject to gradient forces arising from the
dielectric mismatch between colloids and the solvent, and light
pressure \cite{BB04}. Under such forces alone, the colloids have a
preference for regions of highest laser intensity.  In addition, the
(spherical) colloids interact through strong mutual repulsion,
considered to be of a screened Coulomb form \cite{DLVO}
\begin{equation}
\Phi_C = K\sum_{i\neq j} \,\frac{\exp(-\kappa r_{ij})}{\kappa r_{ij}},
\label{eq:yukawa}
\end{equation}
where $r_{ij}$ is the distance between the centres of masses of
macroions $i$ and $j$, $1/\kappa$ is the Debye length \cite{rque2},
and $K$ is an irrelevant prefactor.  To a large extent, colloids in a
given trap experience the repulsion from their $n-1$ ``trap-mates''
only, and interactions with colloids in other traps are of little
relevance to determine the shape that the $n$-mer adopts within a
trap. We therefore assume here that the antagonistic effects of light
interaction and Coulomb repulsion lead to the formation of rigid
composite objects with $n$-fold symmetry (e.g. equilateral triangles
for trimers when there are 3 colloids per trap). This is the basic
assumption of Refs \cite{Agra,Sarlah}, for which the relevant ground
state dimensionless parameters are $\kappa l$, $\kappa d$ and aspect
ratio $\alpha$, where $l$ is the distance between the centres of two
adjacent traps, $d$ is the distance between the centre of the trap and
the centre of one of the colloids forming an $n$-mer (so that in the
dimeric case, $2d$ is the dimer extension, see
Fig. \ref{fig:example}-b)).

The rigidity assumption lumps substrate potential effects into the
length scale $d$ (increasing the well's amplitude decreases $d$). The
position of each $n$-mer is then described by a unique angular
coordinate, while its centre-of-mass always coincides with trap
minimum. Unlike in Ref. \cite{Sarlah}, we do not assume that such
angles are discrete with values dictated by the lattice geometry: it
is indeed of interest to realize that Coulomb repulsion alone is able
to select well defined orientations that do not match any of the
underlying lattice principal direction.  Finding the ground state of a
system of $N$ $n$-mers therefore amounts to minimising the energy
function (\ref{eq:yukawa}) with respect to $N$ angles
\cite{rque3}.

\begin{figure}[!h]
\includegraphics[width=0.66\textwidth]{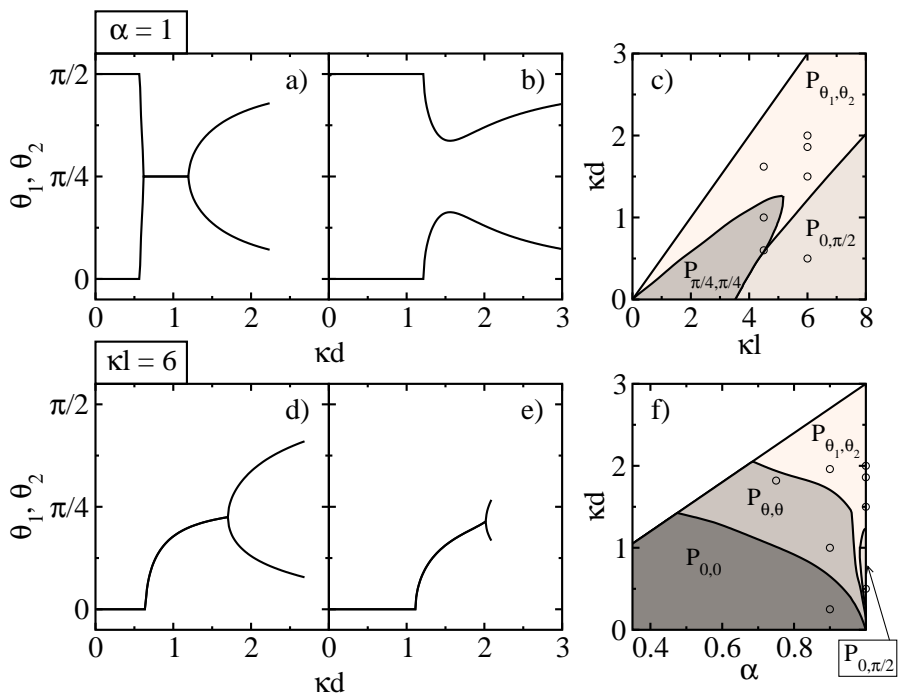}
\caption{Ground state of dimers ($n=2$) on a square 
or rectangular lattice, as obtained from simulated annealing. Graphs
a) b) d) and e) exhibit the dependence of characteristic angles
$\theta_1$ and $\theta_2$ on dimer spacing $d$, lattice constant $l$
and aspect ratio $\alpha$. The first row is for a square lattice
($\alpha=1$) with (a) $\kappa l=4.5$ and (b) $\kappa l=6$. The
corresponding phase diagram is shown in (c). The second row is for
$\kappa l=6$ and (d) $\alpha=0.9$, (e) $\alpha=0.7$, while (f) shows
the phase diagram. As in subsequent figures, the symbols refer to
parameters used for the more complete Monte Carlo simulations of
section \ref{sec:flexible}, which lift the rigidity
assumption. Circles: ``rigid'' simulated annealing and ``flexible''
Monte Carlo simulations agree; triangles for parameters where they do
not agree. Here, no triangles are reported since both approaches yield
similar results, see below.}
\label{fig:dim_square_rigid}
\end{figure}

For the simplest case of dimers ($n=2$) on a square ($\alpha=1$) or
rectangular ($\alpha\neq1$) lattice, we found that in the lowest
energy configuration --obtained by a simulated annealing technique--,
the system adopts a bi-partite structure of the checkerboard type.
The long range orientational order is therefore characterised by two
angles $\theta_1$ and $\theta_2$, which define a phase denoted
$P_{\theta_1,\theta_2}$, see Fig. \ref{fig:example}. Interestingly,
the angles $\theta_1$ and $\theta_2$ are constant in some parameter
range, while they vary continuously in other parameter regions. This
is illustrated in Fig. \ref{fig:dim_square_rigid}. The low $\kappa d$
regions of graphs a) and b) are such that
$(\theta_1,\theta_2)=(0,\pi/2)$, a situation coined
``antiferromagnetic'' in previous studies and illustrated in the lower
right corner of Fig. \ref{fig:example}.  Starting from $\kappa l=6$
with a square lattice (Fig. \ref{fig:dim_square_rigid}-b) and slightly
distorting the lattice into a rectangular one with aspect ratio 0.9,
we observe that the antiferromagnetic $P_{0, \pi/2}$ phase disappears
and turns into a ferromagnetic one $P_{0,0}$, for small enough dimer
extensions (small $\kappa d$), see Figs. \ref{fig:dim_square_rigid}-d)
and -e). For higher $\kappa d$, a tilted ferromagnetic
$P_{\theta,\theta}$ phase appears the most stable, and changes into a
tilted antiferromagnetic $P_{\theta_1,\theta_2}$
(Figs.\ref{fig:dim_square_rigid}-d, -e).  Mild parameter differences
therefore trigger significant orientational changes: for instance, at
$\kappa l = 6$, all phases are antiferromagnetic-like (see graph b))
while at $\kappa l =4.5$ (graph a)) there exists a window around
$\kappa d=1$ with a ferromagnetic ordering. Note also that the upper
part of phase diagram c) corresponds to cases where $d \geq l/2$,
which is ruled out since a given trap cannot extend further than half
the inter-trap distance. Similarly, the forbidden upper region of
diagram f) corresponds to $d \geq \alpha l/2$.

\begin{figure}[!h]
\includegraphics[width=0.66\textwidth]{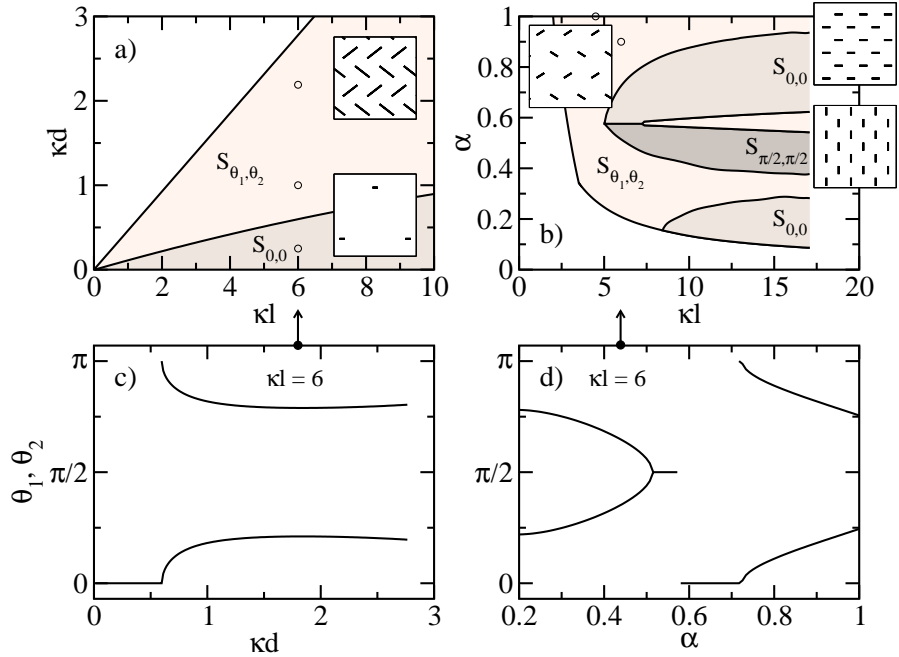}
\caption{Phase diagram for rigid dimers on a triangular lattice for (a) $\alpha=0.9$ and
(b) $\kappa d=1$. The insets show the representative stripe
configurations. The parameter dependence of bipartite angles along the
$\kappa l=6$ line of phase diagrams (a) and (b) is shown in (c) and
(d), respectively. Circles have the same meaning as in
Fig.\ref{fig:dim_square_rigid}.}
\label{fig:dim_triang_rigid}
\end{figure}

\begin{figure}[!h]
\includegraphics[width=0.66\textwidth]{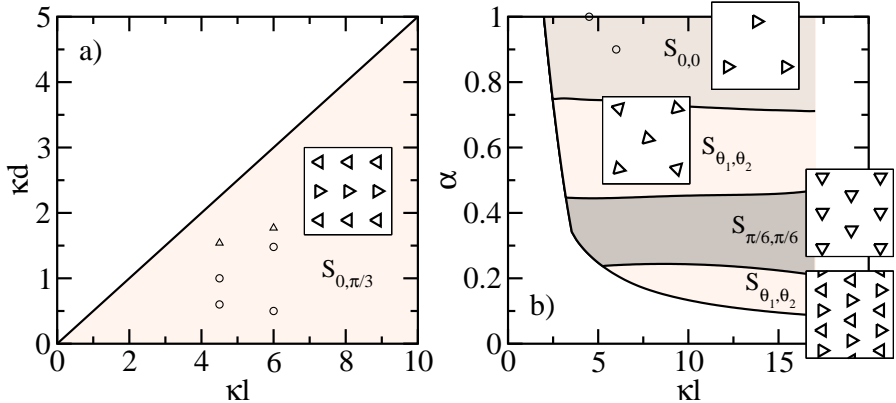}
\caption{Phase diagram of rigid trimers on (a) rectangular lattice for $\alpha=1$ and 
(b) triangular lattice for $\kappa d=1$. The stripe phases are
sketched in the insets, where the centre of masses of the three
colloids gathered in a potential minimum are linked by a thick
line. Here again, circles indicate an agreement between ``rigid'' and
``flexible'' simulations, and triangles report a mismatch.  }
\label{fig:trim_rigid}
\end{figure}

When the lattice geometry is triangular, we have observed the
formation of stripes \cite{Bech2002,RO}, and the corresponding phases
are denoted $S_{\theta_1,\theta_2}$ (the stripes are of the same
bi-partite family as the order reported on the square lattice, since
the checkerboard structure itself is made up of parallel stripes).
The main results are summarised in
Fig. \ref{fig:dim_triang_rigid}. The upper left and lower left regions
of graphs a) and b) respectively are forbidden regions where some
configurations of the rigid trimers in neighbouring traps would lead
to overlaps.  For a given aspect ratio, two ground states are
generically observed : a herringbone order (see e.g. the upper inset
of Fig. \ref{fig:dim_triang_rigid}-a)) and a ferromagnetic one where
dimers align. The sequence of these two phases as parameters are
modified is quite complex, particularly so when the aspect ratio
$\alpha$ is changed. The lattice with maximal symmetry $\alpha=1$
appears singular in that the only phase selected there is the
herringbone one. This qualitatively confirms the results of
\cite{Sarlah}.  The case of trimers on a square or rectangular lattice
is somewhat simpler, with only the stripe $S_{0,\pi/3}$ order observed
(see Fig. \ref{fig:trim_rigid}-a)).  When the trimers are put on the
triangular lattice, several stripe phases are possible, but roughly
speaking, there is a unique type of order for a given aspect ratio,
see Fig. \ref{fig:trim_rigid}-b).

%%%%%%%%%%%%%%%%%%%%%%%%%%%%%%%%%%%%%%%%%%%%%%%%%%%%%%%%%%%%%%%%%%%%%%%%%%%%%%%%%%%%%%%%%%%
\section{Towards a more complete description : effects of flexibility}
\label{sec:flexible}

To test the relevance of the rigidity assumption we now explicitly
take into account the confining potential in the simulations. In order
not to introduce any orientational bias, we consider an isotropic
(harmonic) confining potential $\Phi_L(r)=V_0 (\kappa r)^2$, so that
the total dimensionless energy of colloid $i$ reads:
\begin{equation}
e_i \,=\,E_i/K = 
A(\kappa \delta_i)^2 \,+\, \sum_{j\neq i}  \,\frac{\exp(-\kappa r_{ij})}{\kappa r_{ij}},
\label{eq:withtrap}
\end{equation}
where $\delta_i$ denotes the distance between the centre of mass of
colloid $i$ and the trap centre. The relevant parameters are now
$\kappa l$, $A=V_0/K$ which measures the relative strength of the
light confinement against the Coulomb repulsion and therefore sets the
$n$-mer size denoted $d$ in section \ref{sec:rigid}, and again the
aspect ratio.  We have performed Monte Carlo simulations of this
``flexible'' model, which has $2nN$ degrees of freedom whereas the
``rigid model'' only has $N$ degrees of freedom.

\begin{figure}[!h]
\includegraphics[width=0.52\textwidth]{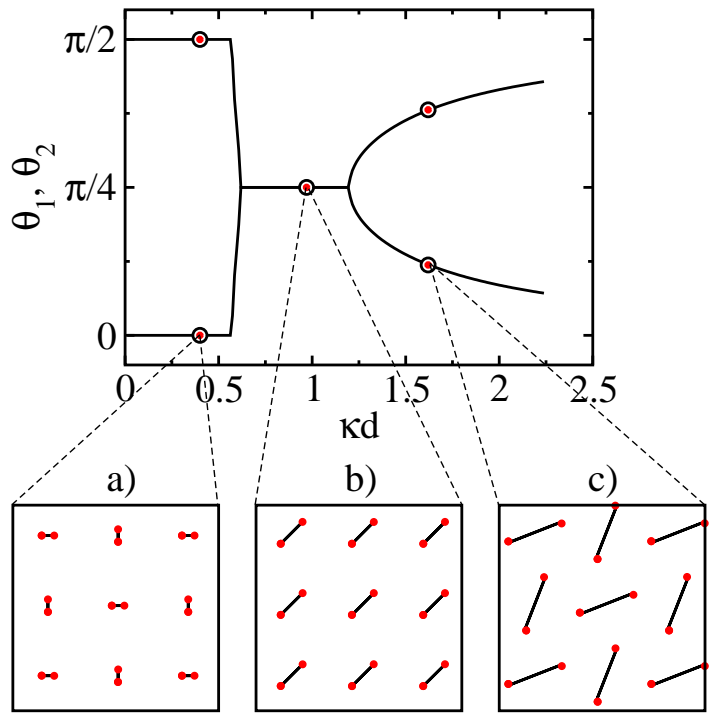}
\caption{Comparison between the ground states for dimers on a square 
substrate, obtained considering the full ``flexible'' model including
explicitly the trapping potential, against the restricted approach
where the $n$-mers are considered as rigid objects. The curve
corresponds to the rigid scenario, while the circles are for its
flexible counterpart. Here, $ \kappa l=4.5$ and (a) $\kappa d=0.4$,
(b) $\kappa d=0.97$, and (c) $\kappa d=1.62$.  }
\label{fig:flexible_dimer_square}
\end{figure}

\begin{figure}[!h]
\includegraphics[width=0.6\textwidth]{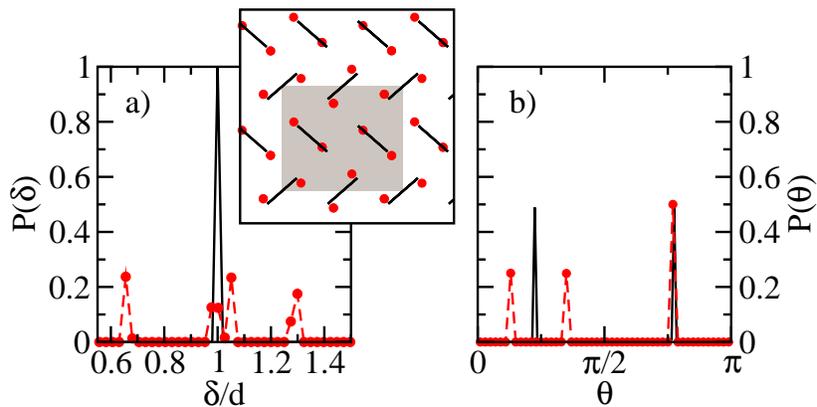}
\caption{Emergence of tetrapartite ground-state configurations for flexible
dimers on a triangular lattice with $\kappa l=6$, $\kappa d=1.94$,
$\alpha=1$. The ``rigid'' configuration is shown for comparison with
dimers indicated by lines, while the discs materialise the colloids in
the flexible ground state. Graph a) shows the probability distribution
function of colloid position in the trap ($\delta =0$ for the trap
centre). Graph b) is for the distribution function of the angle
between a reference direction and the line joining the trap centre to
a given colloid.  As for the inset showing the configuration, the
flexible data are shown with discs and the rigid ones are shown by the
continuous line.}
\label{fig:flexible_dimer_triangular}
\end{figure}

\begin{figure}[!h]
\includegraphics[width=0.6\textwidth]{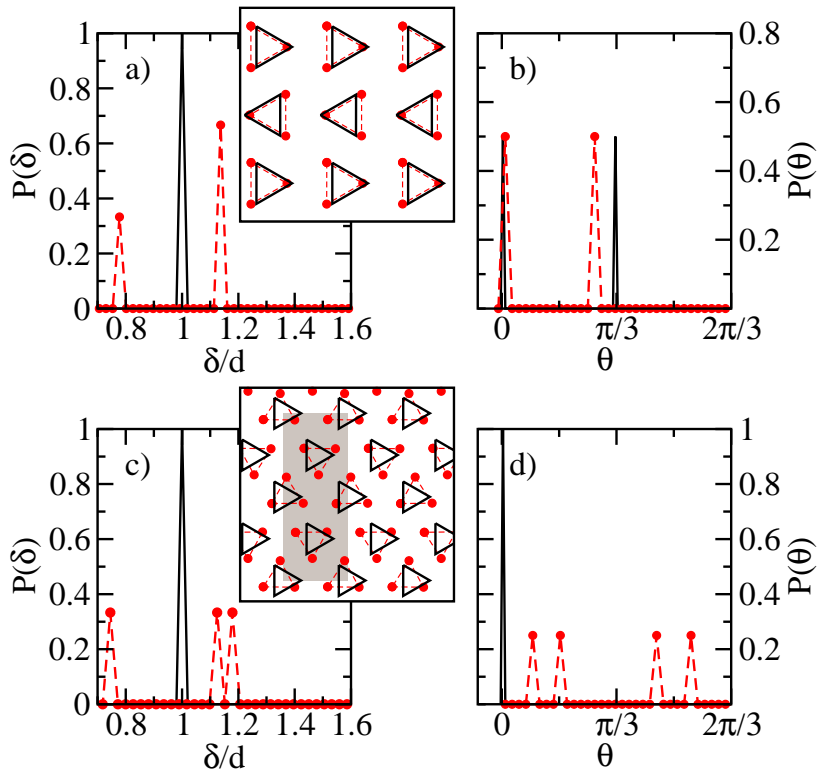}
\caption{Same as Figure \ref{fig:flexible_dimer_triangular} for trimers
on a rectangular lattice (graphs a) and b) for which $\kappa d=1.88$,
$\kappa l=6$ and $\alpha=0.9$), and for trimers on a triangular
lattice (graphs c) and d) for which $\kappa d=1.66$, $\kappa l=6$ and
$\alpha=0.75$). }
\label{fig:flexible_trimer}
\end{figure}

The results are shown in Figs. \ref{fig:flexible_dimer_square},
\ref{fig:flexible_dimer_triangular} and \ref{fig:flexible_trimer}, and depend
on the particular situation. The comparison flexible/rigid was
performed by first implementing simulated annealing for the flexible
model, measuring the resulting $n$-mer size, and using the
corresponding value of $d$ in a rigid-model simulation. For dimers on
the square/rectangular lattice, we always found an excellent agreement
between both routes, see Fig. \ref{fig:flexible_dimer_square} where
three typical bipartite orders are shown in the insets. On the
triangular lattice, we found that the qualitative features put forward
in section \ref{sec:rigid} remain correct (see e.g. the configuration
shown in the inset of Figure \ref{fig:flexible_dimer_triangular}),
while on closer inspection, some differences arise. Indeed, the
histograms of distances to trap centre $\delta$ and tilt angle
$\theta$ in Figure \ref{fig:flexible_dimer_triangular} clearly reveals
the mismatch between the rigid and flexible ground states that are
respectively of bipartite and of quadripartite type (the inset shows
the new unit cell by a shaded area).  Correlating the histograms to
the snapshot of the inset, it appears that the top row of the inset
corresponds to the most intense peak in the $P(\theta)$ distribution,
and to the two extreme peaks in $P(\delta)$ at $\delta\simeq 0.65 d$
and $1.35 d$. Conversely, the second row from the top has colloids
contributing to the two smaller peaks in $P(\theta)$ that weight half
the previous one, and contributing to the peaks at $\delta \simeq
d$. It can be seen that the average $\delta$ is $d$, as it should from
the definition of $d$ in the flexible case.

Figure \ref{fig:flexible_trimer} shows similar results for trimers on
both rectangular and triangular lattices. On the rectangular lattice,
the main difference with the rigid case (where the trimer centre is
imposed to coincide with the trap centre), lies in an off-centre
shift, see the inset of graph a). Graphs a) and b) corroborate and
quantify this visual observation. On the triangular lattice, the
effect is more spectacular. While the rigid model leads to an order
where all trimers align in the same direction ($S_{0,0}$ fashion), due
account of the internal flexibility of the trimers yields an order
that is visually reminiscent of a stripe phase $S_{\pi/6,\pi/2}$, but
that is in reality more complex.  This can be appreciated by comparing
the configurations in the first and third rows from the top of the
inset in graph c) (or equivalently second and fourth): an off-centre
shift of the triangles may be observed, and substantiated by the
histograms of graphs c) and d). Here again, the order selected is of
tetrapartite form, with a unit cell shown by the shaded area, and not
bipartite as found in section \ref{sec:rigid}.

%%%%%%%%%%%%%%%%%%%%%%%%%%%%%%%%%%%%%%%%%%%%%%%%%%%%%%%%%%%%%%%%%%%%%%%%%%%%%%%%%%%%%%%%%%%
\section{On the relevance of different approximation schemes}
\label{sec:approx}

In the previous section, we have shown that assuming the trapped
colloids to form a rigid composite object may be incorrect in some
cases.  Here, we address the applicability of two other classes of
simplification of the original model provided by the energy function
(\ref{eq:withtrap}).

\subsection{The nearest neighbour interactions}

\begin{figure}[!h]
\includegraphics[width=0.6\textwidth]{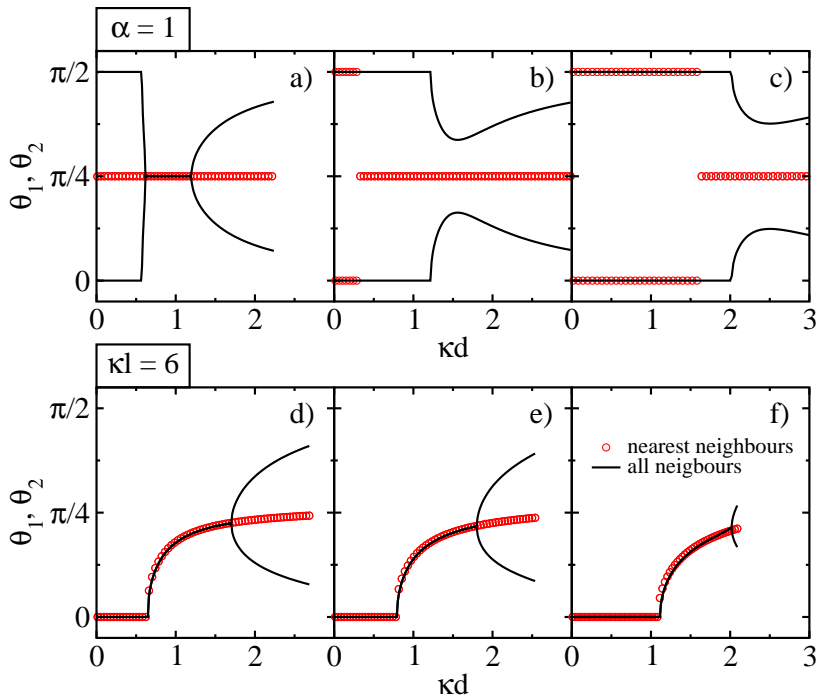}
\caption{Comparison of the nearest neighbour assumption (shown with symbols)
and the correct result (shown with solid lines) for dimers on a
rectangular lattice. The top row is for a square geometry
($\alpha=1$), with a) $\kappa l=4.5$, b) $\kappa l=6$ and c) $\kappa
l=8$. Second row is for $\kappa l=6$, tuning the aspect ratio: d)
$\alpha=0.9$, e) $\alpha=0.85$ and f) $\alpha=0.7$.}
\label{fig:nncomparisonRect}
\end{figure}

\begin{figure}
\includegraphics[width=0.6\textwidth]{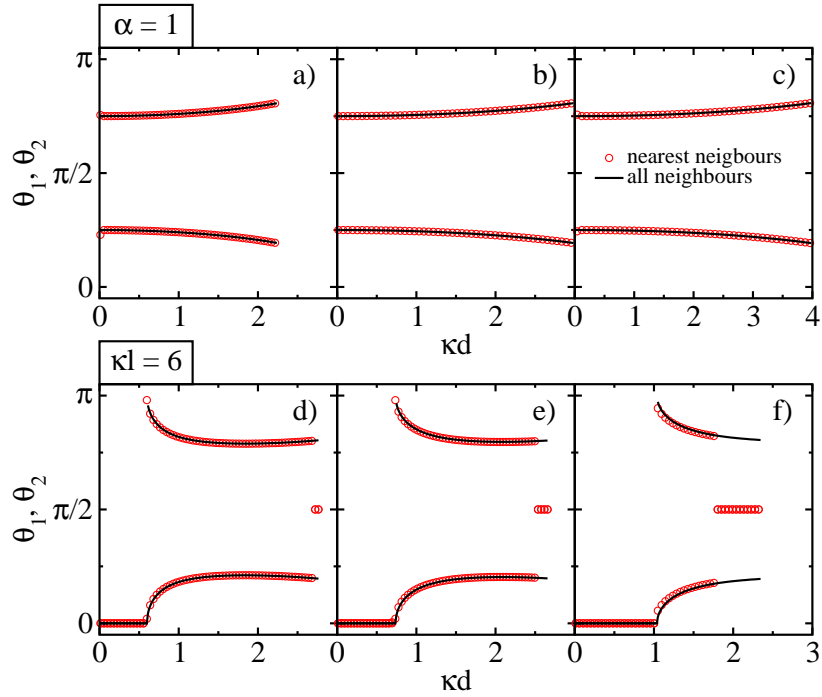}
\caption{Comparison of the nearest neighbour assumption (shown with symbols)
and the correct result (shown with solid lines) for dimers on a
triangular lattice. The parameters are the same as in
Fig. \ref{fig:nncomparisonRect}.}
\label{fig:nncomparisonTrian}
\end{figure}

\begin{figure}[!h]
\includegraphics[width=0.6\textwidth]{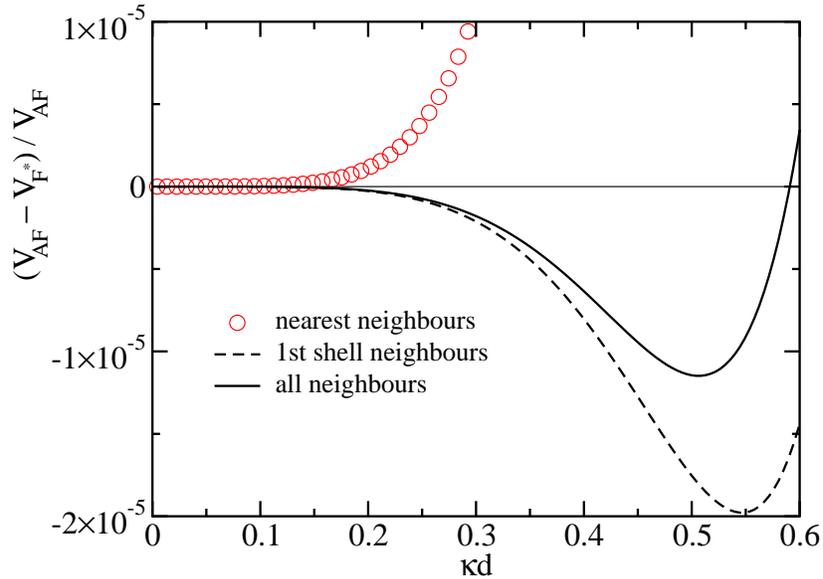}
\caption{Comparison of the energies of the $P_{0,\pi/2}$ phase (denoted AF) and
the $P_{\pi/4,\pi/4}$ phase (denoted F$^*$) for dimers on a square
lattice ($\alpha=1$) with $\kappa l=4.5$ and small $\kappa d$. This
parameter range corresponds to the small $\kappa d$ region of
Fig. \ref{fig:nncomparisonRect}-a). The symbols are for nearest
neighbour interactions only, the dashed line adds the next shell of
nearest neighbours (the four corner sites in the square), while the
solid curve is for the case where all neighbours in all traps are
considered.  Due to the extremely small energy differences
$V_{AF}-V_{F^*}$ summation over more neighbour shells is required and
the nearest neighbour approximation, although intuitively expected to
be valid, fails.}
\label{fig:comp_en_afvsf*}
\end{figure}

The first simplification, considered in both Refs. \cite{Agra} and
\cite{Sarlah} consists in restricting colloid interactions to partners
in nearest neighbour traps. Given the exponential character of
screened Coulombic law (\ref{eq:yukawa}), this seems an {\it a priori}
reasonable assumption, provided both the distance between adjacent
traps ($l$) and the closest distance between colloids ($l-2d$) are
large enough compared to Debye length $1/\kappa$. The comparison
between the two approaches is displayed in
Fig. \ref{fig:nncomparisonRect} for dimers on a square or rectangular
lattice, and on Fig. \ref{fig:nncomparisonTrian} for dimers on a
triangular lattice. We consider only the rigid dimers, which has been
shown to be sufficient in section \ref{sec:flexible}, see
Fig. \ref{fig:flexible_dimer_square}). The above argument leads to
believe that for large enough $\kappa l$ and small enough $\kappa d$,
interactions beyond the nearest neighbours should be immaterial. This
can be observed in Fig. \ref{fig:nncomparisonRect}, where the angles
$\theta_1$ and $\theta_2$ (those of the bipartite checkerboard
structure) are plotted against $\kappa d$ for six different ($\kappa
l,\alpha$) combinations. The nearest neighbours approximation
generally works well when $\kappa d \ll \kappa l$ with, however, the
surprise that at $\kappa l=4.5$ and small $\kappa d$, the nearest
neighbour route is quantitatively wrong, leading to a tilted
ferromagnetic phase $P_{\pi/4,\pi/4}$ instead the antiferromagnetic
$P_{0,\pi/2}$ (see Fig. \ref{fig:nncomparisonRect}a)). A comparison
between the energies of the different phases is performed in
Fig. \ref{fig:comp_en_afvsf*}. Surprisingly, the relative energy
difference between ferromagnetic and antiferromagnetic phases is
minute (see the $y$-scale) and explains why including second nearest
neighbours is essential to get the correct phase behaviour. The dashed
line in Fig. \ref{fig:comp_en_afvsf*} is obtained by summing over the
four nearest neighbour traps and over the four next-nearest corner
traps in the rectangular array. Restricting to nearest neighbours
leads to the prevalence of the tilted ferromagnetic (symbols), which
is incorrect, although energetically very close to the
antiferromagnet. When $\kappa l$ is increased, this quasi degeneracy
disappears, and including next nearest neighbours in the analysis
becomes irrelevant.

In the case of the triangular lattice, the nearest neighbour
approximation is more efficient than on the rectangular lattice, as
can be seen in Fig. \ref{fig:nncomparisonTrian}. The reason for this
is most likely that in the triangular lattice where there are six
nearest neighbour sites, the next-nearest neighbour distance is
$\sqrt{3}$ times larger than the nearest neighbour one. On a square
lattice on the other hand, there are four nearest neighbour sites and
the next-nearest distance is only a factor $\sqrt{2}$ larger. The
situation is thus less favourable for nearest neighbour truncation.

\subsection{The large-distance approximation}

The second simplification enforced in \cite{Agra} with the purpose to
allow for the construction of a tractable Hamiltonian, which can be
mapped onto a spin model, amounts --within the rigid scenario-- to
considering the leading order term only in the large distance
expansion of the interaction potential. The reason for doing so is
again to focus on the large $\kappa l$ case, with the technical bonus
that then, interactions between 2 $n$-mers in different traps may be
written in a factorized way. To be specific, the large distance
potential of interaction between two dimers labelled 1 and 2, of size
$2d$, with centre-of-mass/centre-of-mass separation $\mathbf{r}$ reads
\begin{equation}
V_{12} \,=\, \cosh[\kappa d  \cos(\theta_{12})] \, \cosh[\kappa d  \cos(\theta_{21})] \,
\frac{e^{-\kappa r}}{\kappa r},
\label{eq:cosh}
\end{equation}
where $r=|\mathbf{r}|$ and $\theta_{ij}$ is the angle between vector
$\mathbf{r}$ and the direction defined by dimer $i$. Similar
considerations prove generically fruitful to discuss interactions
between anisotropic colloids in the low density regime
\cite{JPCM1,EPJE}.
\begin{figure}[!h]
\includegraphics[width=0.6\textwidth]{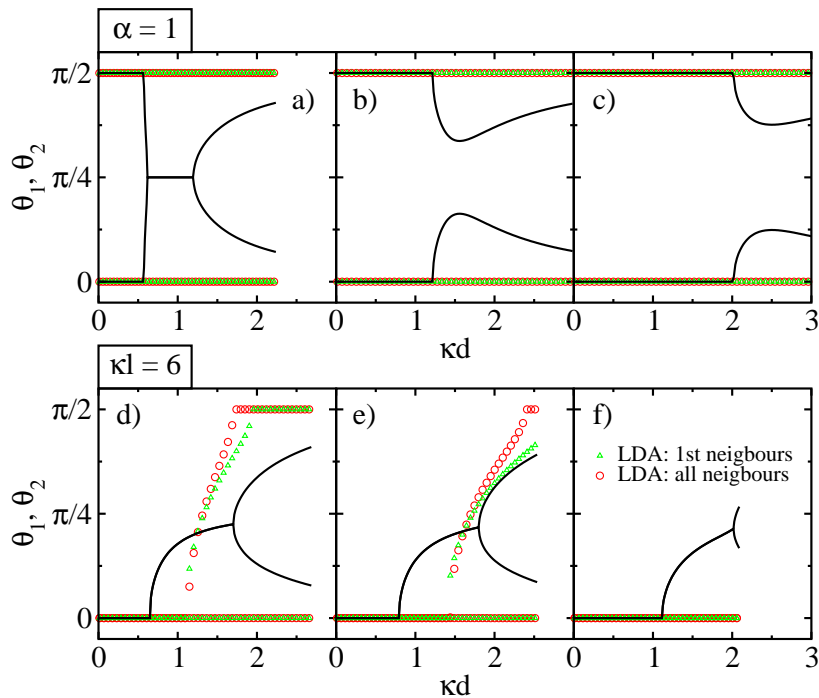}
\caption{Effect of considering the large distance approximation (LDA) of the
interaction potential between dimers (on rectangular lattices) on the
bipartite angular ordering. The parameters are the same as in
Fig. \ref{fig:nncomparisonRect}.  The truncated potential results with
expression (\ref{eq:cosh}) are shown with the symbols: triangles
restricting to nearest neighbours, and circles including all
neighbours.  The results are compared to the ``all neighbours''
simulation of the rigid dimers (solid lines).}
\label{fig:coshcomparison}
\end{figure}
It can be seen in Figure \ref{fig:coshcomparison} that for low $\kappa
d$, the corresponding predictions fare favourably against the results
of the full expression (\ref{eq:yukawa}).  However, the truncated
approach always predicts the antiferromagnetic phase in the square
case \cite{Agra}, and misses the tilted antiferromagnet (see the top
row).  On a rectangular lattice, it correctly captures the transition
from a ferromagnet $P_{0,0}$ to tilted ferromagnet, but overestimates
the threshold (see the bottom row), to such an extent that it can
exceed $\alpha l/2$, the maximum allowed value of $d$ (see graph f)).
The failure at larger $\kappa d$ has a different origin from the one
observed in Fig. \ref{fig:nncomparisonRect}, and could have been
anticipated by computing the sub-dominant terms in expression
(\ref{eq:cosh}). For the sake of clarity, we focus here on the
electric potential created by a dimer, a simpler but related object
than the dimer-dimer potential considered in (\ref{eq:cosh}); the sum
of the two screened Coulomb terms associated to each colloid may be
written
\begin{equation}
 V(\mathbf{r}) \, \simeq \,\cosh[\kappa d  \cos(\theta)] \, \exp\left(\frac{(\kappa d)^2}{2\kappa r}(\cos^2\theta-1) \right)
\frac{e^{-\kappa r}}{\kappa r},
\end{equation}
where $\theta$, defined as above, is the angle under which the dimer
is seen from a distance $r$. Neglecting the second factor on the right
hand side (which leads to Eq. (\ref{eq:cosh}) for the interaction) is
only justified provided $(\kappa d)^2 < \kappa l$.  Finally, we note
that although the parameters in Figs. \ref{fig:nncomparisonRect} and
\ref{fig:coshcomparison} are exactly the same, truncating colloid 
interactions to nearest neighbour traps only makes little difference
if the potential is of the form (\ref{eq:cosh}). This is at variance
with what can be observed in Fig. \ref{fig:nncomparisonRect}.
We are back here to the message conveyed by Figure \ref{fig:comp_en_afvsf*}, that minor
modifications of the original problem may significantly alter the preferred phase.

\subsection{The discrete angles approximation}

In reference \cite{Sarlah}, an effective Potts-like Hamiltonian was
constructed by assuming rigid molecules of fixed sizes placed on the
lattice points, restricting the orientations $\theta_i$ of the
molecules to discrete values compatible with the lattice symmetry and
by considering the nearest neighbour interactions -- Yukawa
interactions between colloids in neighbouring traps. Since the phase
behaviour studied in \cite{Sarlah} pertains to triangular substrate
potentials, the nearest neighbour approximation seems well justified,
see Fig. \ref{fig:nncomparisonTrian}. The restriction to discrete
angles, which rests on the remark that the confinement potential
exhibits some preferred directions, however deserves a more careful
discussion.

For concreteness, we consider in the triangular geometry case, a
confinement potential of the same form as in Ref. \cite{RO}:

\begin{eqnarray}
\phi_L = const \, -\,  \Biggl [ \cos\left(2\pi\frac{x-y/\sqrt{3}}{l}\right )\,+
\,\cos\left (4\pi \frac{y}{l\sqrt{3}}\right )\,+\,
\cos\left (2\pi\frac{x+y/\sqrt{3}}{l}\right )\Biggr ]\;.
\label{anisoV}
\end{eqnarray} 

The iso-$\phi_L$ lines are isotropic (circular) in the vicinity of the
minima, with an anisotropy that increases with increasing energy
$\phi_L$. As a consequence, the confinement potential anisotropy is
all the more important as the $n$-mer is more extended (large $d$),
but this parameter range corresponds to a situation of strong
Coulombic repulsion where details of the light potential may not
matter. It is therefore not straightforward to anticipate the relative
ranges of applicability of the discrete angle approximation of
\cite{Sarlah}, and of the isotropic confinement potential approach
followed in this paper, which leads to
Figs. \ref{fig:example}-\ref{fig:coshcomparison}.

\begin{figure}[!h]
\includegraphics[width=0.72\textwidth]{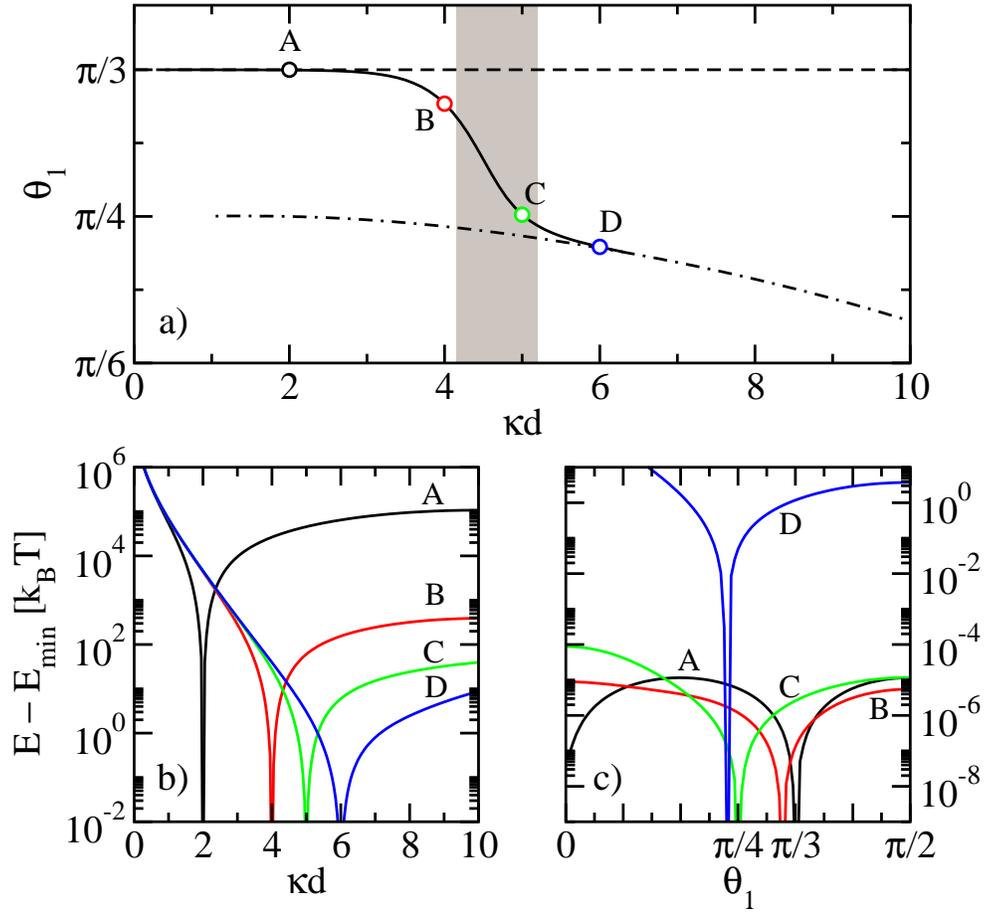}
\caption{a) The optimal orientation $\theta$ of a dimer in the triangular light
potential obtained through minimization of the total energy
(confinement+Coulombic). The dot-dashed line is for an isotropic
parabolic light potential where only the Yukawa part of the energy
determines the angle $\theta$. The upper dashed line is the constant
value $\theta=\pi/3$ compatible with the lattice symmetry of the light
potential. The solid line is for the case of anisotropic confining
potential (\ref{anisoV}) where orientations favoured by Yukawa
interactions and those favoured by the lattice symmetry, compete. The
shaded area represents the parameter regime of the experiments
\cite{Bech2002}. b) Radial cut through the total energy landscape for
different points in graph a) corresponding to different values of the
ratio $A$. The depth of the minimum represents the binding strength of
the dimer in radial direction. The sizes of the molecule are
determined from the location of this minimum. c) Angular cut through
the total energy landscape for the same points as in graph b). The
depth of a minimum represents the angular binding strength of the
dimer. This angular binding energy is of the order of a few $k_B T$ at
large $\kappa d$ (point D) and very small for smaller molecules
(curves A, B, C). The total energy $E=Ke$ in b) and c) is obtained
assuming $K=10^5 k_B T$.}
\label{fig:Angles}
\end{figure}

To answer this question, we have minimised the total energy
$e=A\sum_i\phi_L+\phi_C$, with respect to the angular orientation of
the dimer. Such an approach takes due account of the interplay between
the realistic potential (\ref{anisoV}) and Coulombic interactions.
The ground state results are shown in Fig. \ref{fig:Angles}, for the
same parameters as in the experiments of Ref \cite{Bech2002}:
$\alpha=1$, $\kappa l= 20$. The dimers are assumed to form the
herringbone structure $S_{\theta_1,\theta_2}$, where due to the
symmetry, we may assume $\theta_2=\pi-\theta_1$ (see
Fig. \ref{fig:dim_triang_rigid}). For a given ratio $A$ between the
light confinement and the strength of the Yukawa interaction, the
total energy $e$ has been minimised with respect to the radial
($\kappa d$) and angular ($\theta_1$) positions of the colloid inside
the trap. The resulting angles $\theta_1$ are plotted against $\kappa
d$ in Fig.\ref{fig:Angles}(a), see the solid line. For comparison, the
lower dot-dashed curve is the result obtained with the isotropic
potential (angles selected by the Yukawa interactions). The upper
dashed curve is the constant angle $\theta_1=\pi/3$ assumed in
\cite{Sarlah}. It appears that the angles are compatible with the pure
Yukawa prediction at large values of $\kappa d$. In essence, although
the confinement potential is more anisotropic at large $d$
(corresponding to small $A$), the Yukawa term nevertheless dominates
the total energy so that the isotropic potential approach becomes
correct.  Conversely, the discrete angle approach appears more
relevant for small colloidal molecules.

In Fig. \ref{fig:Angles}(b) we show the radial dependence of the total
energy, fixing $\theta$ to its optimal value. The total energy $E=Ke$
is obtained assuming the experimentally realistic value $K=10^5 k_B
T$, the value used also in \cite{Sarlah}. Larger molecules have weaker
radial binding energies and are therefore more prone to fluctuate at
finite temperature.  In Fig.\ref{fig:Angles}(c), we plot the angular
dependence of the total energy at fixed $\kappa d$. The depth of the
minimum can be understood as the angular binding energy. This binding
energy is of several $k_B T$ at large $\kappa d$, while it is very
small at small $\kappa d$, in the $\theta=\pi/3$ regime
\cite{rque101}. The assumption of discrete angles therefore fails at
large $\kappa d$ (weak light potentials) and leads to the correct
results only at small $\kappa d$, where, however, due to the shallow
binding in the angular direction, there is no {\sl a-priori} reason to
justify it but for ground state properties.

The parameter regime of the experiments \cite{Bech2002} is marked as a
shaded area in Fig. \ref{fig:Angles}(a). Interestingly, the
experimental situation is somewhat intermediate between both limiting
regimes and neither of the approaches seems to be perfectly
appropriate. However, other realizations of the light confinement are
possible, like creating isotropic optical point traps instead of using
the interference patterns as in \cite{Bech2002}. For such an
experiment, the present approach -- the lower curve in
Fig.\ref{fig:Angles}(a), corresponding to the isotropic trap -- would
be relevant.

Finally, we also note that numerical simulations of colloidal
molecules in the anisotropic light confinement (\ref{anisoV}) have
been performed in \cite{RO}. The parameters studied there ($\kappa
l=2$) are nevertheless quite distant from those of the experiments
\cite{Bech2002} where $\kappa l\approx 20$. In an aqueous solution,
$\kappa l=2$ cannot be achieved with 3 $\mu m$ size colloids as used
in \cite{Bech2002}, since the inverse Debye length $\kappa$ is bounded
from below by the solvent dissociation. Moreover, in the regime
$\kappa l \approx 2$, the pair-wise additivity assumption leading to
the Yukawa potential is questionable, and many-body effects could play
an important role.

%%%%%%%%%%%%%%%%%%%%%%%%%%%%%%%%%%%%%%%%%%%%%%%%%%%%%%%%%%%%%%%%%%%%%%%%%%%%%%%%%%%%%%%%%%%
\section{Conclusion}
\label{sec:concl}

We have provided a numerical analysis of the type of long range
orientational orders selected in so-called colloidal molecular
crystals, where a given integer number $n$ of colloids is trapped in
each potential minimum of a light lattice.  We investigated the cases
of dimers ($n=2$) and trimers ($n=3$) on both square and triangular
lattices, together with deformed geometries obtained when a given
direction of the original lattice is expanded or shrunk by a factor
$\alpha$. The sequence of phases reported is rich.  For triangular
lattices mostly, and although we were only interested in the lowest
energy configuration, we have uncovered the relevance of the
flexibility of the $n$-mers in a given trap, whereas previous
approaches envisioned those composite objects as a rigid entity. In
most cases, the qualitative predictions derived within the rigid
picture appear qualitatively correct though. We have also addressed
the adequacy of truncating interactions to colloids located in nearest
neighbour traps, and found that in some cases, the quasi degeneracy
between states of different orientational orders makes second nearest
neighbours and possibly more remote shells relevant.

We emphasise that the pinning potential we considered is of isotropic
form, and therefore does not favour any orientation of the trapped
$n$-mers. On the other hand, it was assumed in Ref. \cite{Sarlah} that
the pinning potential itself leads to a discrete set of possible
orientations, which were supposed to match the lattice symmetry. In
particular, such an assumption would only be compatible with the
ferromagnetic $P_{0,0}$ and antiferromagnetic $P_{0,\pi/2}$ phases for
dimers on the square lattice.  Our investigation shows that several
other tilted phases exist, with an angular selection due to Coulombic
repulsive interaction alone. Instead of a discrete set of predefined
angles we find that the angular coordinates change continuously with
parameters. We have explored in detail the limits of validity of both
assumptions and have concluded that working with discrete angles can
only be justified if the light confinement is very strong or when
colloids repel weakly. We have shown however that the angular
confinement is very weak in this regime.

{\bf Acknowledgements: } We would like to thank Fr\'ed\'eric van
Wijland for useful discussions and P. Ziherl for carefully reading the
manuscript. The work was supported by the Slovenian Research Agency
through grants P1-0055 and Z1-9303. We acknowledge the support of the
bilateral program Proteus supported by the slovenian Research Agency
and the french Minist\`ere des Affaires Etrang\`eres et Europ\'eennes.


\begin{thebibliography}{99}
\bibitem{rque2}
While colloidal positional degrees of freedom are limited to a  2D plane, microions 
and solvent molecule are not subject to confinement, which results in a traditional
three-dimensional-like screened Coulomb interaction between macroions.
\bibitem{rque3}
Strictly speaking, expression (\ref{eq:yukawa}) is not an energy but a free energy
since it includes an entropic contribution from the microions; we also
stress that the ground state we consider are reached at constant $\kappa$
while the latter is a temperature dependent quantity; this is in line with the 
numerical approach of \cite{RO}, and relevant to experiments where the
aforementioned ground state may be approached by tuning the pinning amplitude
at constant temperature.
\end{thebibliography}
\end{document}